\documentclass[11pt]{article}
\usepackage{bnaic}

\usepackage[ruled,vlined]{algorithm2e}
\usepackage{graphicx}
\usepackage{tikz}
\usepackage{caption}
\usepackage{array}
\usepackage{adjustbox}
\usepackage{makeidx}
\usepackage{pdflscape}
\usepackage{hyperref}
\usepackage{threeparttable}
\usepackage{rotating}
\usepackage{subcaption}
\usepackage{amsmath}
\usepackage{amsfonts}
\usepackage{amssymb}
\usepackage{listings} 
\usepackage{xcolor}   
\usepackage{float}
\usepackage{tablefootnote}
\usepackage{titlesec}
\titleformat{\section}[block]{\normalfont\Large\bfseries}{}{0pt}{} 
\titleformat{\subsection}[block]{\normalfont\large\bfseries}{}{0pt}{} 
\titleformat{\subsubsection}[block]{\normalfont\normalsize\bfseries}{}{0pt}{} 

\title{\textbf{Evaluating Front-end \& Back-end of Human Automation Interaction Applications \\ $\Delta$-EVAL \\ A Hypothetical Benchmark}}
\author{Gonçalo Hora de Carvalho \\(goncalo@iiim.is)}
\date{}

\begin{document}

\begin{figure}
    \centering
\end{figure}

\maketitle

\begin{abstract}
\noindent
Human Factors, Cognitive Engineering, and Human-Automation Interaction (HAI) form a trifecta, where users and technological systems of ever increasing autonomous control occupy a centre position. But with great autonomy comes great responsibility. It is in this context that we propose metrics and a benchmark framework based on known regimes in Artificial Intelligence (AI). A benchmark is a set of tests and metrics or measurements conducted on those tests or tasks. We hypothesise about possible tasks designed to assess operator-system interactions and both the front-end and back-end components of HAI applications. Here, front-end pertains to the user interface and direct interactions the user has with a system, while the back-end is composed of the underlying processes and mechanisms that support the front-end experience. By evaluating HAI systems through the proposed metrics, based on Cognitive Engineering studies of judgment and prediction, we attempt to unify many known taxonomies and design guidelines for HAI systems in a benchmark. This is facilitated by providing a structured approach to quantifying the efficacy and reliability of these systems in a formal way inspired by the recent fast developments in AI benchmarking techniques, thus, we attempt to guide designing principles towards a testable benchmark capable of reproducible results that is future-proof, general, and insightful both in the cognitive and technological stacks of any HAI application.
\end{abstract}

\section{Introduction}
Benchmarks are sets of tests or tasks and their associated metrics used in evaluating a system comparatively. Such programs help discover the relative state of the art by comparing systems with each-other relative to some measurement(s) outcomes. They are important, for example, for researchers that are interested in measuring and comparing algorithmic or system performance, in tracking progress in the field, ensuring reproducibility, driving innovation, and providing objective comparisons between competing theories. Scientifically, they may involve standardised datasets, measurements (used interchangeably with 'metrics'), clear task definitions, established rules, and baseline results for fair and consistent assessments \cite{lm_evaluation_harness, rajpurkar2016squad, wang2018glue, srivastava2023imitation, lintang_sutawika_2023_10256836}. For the rest of this text, we will focus on the specific case of benchmarking in Artificial Intelligence (AI) and how the philosophy and explicit techniques can be useful for HAI application research and development. 

A common goal of Cognitive Engineering, Human Factors, and Human Automation Interaction (HAI) is the construction of contextually appropriate systems that enable human operators to perform tasks assisted by or enabled by digital systems \cite{chapter4}. A problem they share in designing any system is its evaluation, often requiring in-depth meta-analysis which can quickly become a problem in its own right \cite{lm_evaluation_harness, srivastava2023imitation}. This difficulty arises because systems in Cognitive Engineering, Human Factors, and HAI are highly context-dependent and interact with complex human behaviors, making standardisation hard to achieve \cite{chapter4}.

Moreover, the effectiveness of such systems is often measured not only by quantitative outputs but also by qualitative factors such as user satisfaction and cognitive load \cite{chapter4, Automation_failure, effects_imperfect, karanikas}. Traditional benchmarks often fail to capture the full spectrum of human-centered metrics, focusing instead on technical performance alone or in the surrounding environment. For example, in the study by Karanikas et al., the researchers compiled an overview of existing methods for measuring safety performance, showing that these typically involve direct numerical data such as accident rates, hazard reports, and incident reports, along with survey results and cost-benefit analyses. The authors conclude that these methods may help identify what safety issues are present and where they occur within an organization, however, they do not explain why these issues happen and how to effectively address them \cite{karanikas}. We contend that in order to further our understanding in measuring the quality of a system design, we must also address the loci of action - the human operator.

The problem of defining performance metrics that encompass both the technical and human aspects of system functionality further complicates the evaluation process \cite{endsley, stowers}. Thus, there remains a significant challenge in developing benchmarks that effectively reflect the complexities and multi-dimensional nature of systems designed for human operators to interact with such as the ones relevant to Cognitive Engineering, Human Factors, and Human-Automation Interaction systems \cite{chapter4}. This limitation is what drives the current work and development of $\Delta$-EVAL: the need for a more holistic approach to assess both human and machine considerations in one singular benchmark.

\section{Background}

One step towards such a benchmark can be found in Stowers et al., but while their proposed framework for assessing human-machine systems is a good review of previous attempts, it is still quite vague in that it does not offer concrete metrics (mathematically defined) - see Table \ref{tab:stowersmetrics}. The criteria for measuring safety, performance, and human processes—particularly cognitive states like trust and situation awareness are not clearly defined. For instance, the framework outlines the measurement of "trust in automation" and "situation awareness" without specifying the formal mathematical definition, tools, or methods to reliably assess these complex psychological constructs. Similarly, the metrics for system inputs such as "level of automation" and "adaptiveness" are mentioned without clear guidelines on how to quantify these characteristics. This lack of specificity and formalisation hinders the practical utility of the framework \cite{stowers}.

\begin{table}[H]
\centering
\begin{tabular}{|l|l|}
\hline
\textbf{Metric Category} & \textbf{Specific Metrics} \\ \hline
Safety                  & Absence of catastrophic failures, accident rates \\
Performance             & Goal completion, efficiency, effectiveness \\
Human Processes         & Trust in automation, reliance, situation awareness \\
Inputs                  & Human expertise, machine reliability, environmental factors \\ \hline
\end{tabular}
\caption{Summary of Proposed Metrics in Stowers et al. (2017) \cite{stowers}.}
\label{tab:stowersmetrics}
\end{table}

In contrast with Stowers et al.'s benchmark which attempts to measure a particular facet of human operators \cite{stowers}, we aim to propose a benchmark that unifies Cognitive Engineering, Human Factors, and HAI. Cognitive Engineering primarily studies the cognitive processes involved in system interactions. It employs principles from cognitive science to design and evaluate systems in a way that aligns with how people think, perceive, remember, and act. Attention and judgment are at the core of these cognitive processes \cite{chapter4}. Attention pertains to how individuals allocate their cognitive resources to process specific pieces of information in an environment filled with stimuli \cite{lee2013oxford}. Judgement involves the evaluation or estimation of situations, often leading to decision-making, based on the information perceived \cite{lee2013oxford}.
Human Factors, on the other hand, classically focuses on understanding and designing systems that are congruent with human capabilities and limitations. These encompass physical, cognitive, and organizational aspects \cite{wickens2004introduction}. 

Human operators interface with computational systems which become increasingly automated. In order to guarantee their reliable performance while tailoring them to accommodate human cognition as per cognitive engineering and human factors, design and evaluation procedures become increasingly complex, difficult, and important. This is because they must seamlessly integrate with human operators' cognitive processes and capabilities, ensuring efficient collaboration and minimizing errors, thereby maximizing overall system effectiveness and safety \cite{chapter4, sheridan2005human}. It is in this intersection between human factors, cognitive engineering, and automated systems that Human-Automation Interaction (HAI) emerges \cite{sheridan2005human}. HAI mainly concerns information processing that can be broadly categorized into two processes: front-end and back-end \cite{chapter4}.
When discussing front-end and back-end in the context of these disciplines, specifically in HAI, they can be seen as the two ends of the interaction spectrum. The front-end is the interface or point of contact between the human and the system or machine, where immediate interactions (input/output), perceptions, and judgments occur. It encompasses everything the user directly interacts with or perceives, from graphical interfaces to tactile feedback \cite{chapter4}. 

The back-end, conversely, refers to the underlying processes, algorithms, and mechanisms that support and drive system control as well as the front-end experience. While users might not directly interact with the back-end, its efficiency and reliability significantly influence their overall experience and trust in the system \cite{chapter4, Automation_failure, effects_imperfect}.

When benchmarking HAI in the context of front-end and back-end, human factors is relevant in ensuring that systems have been designed with human needs in mind. These needs can be encoded as evaluation metrics and measured. HAI narrows these down to contexts where humans interact with automated systems and make decisions based on these interactions. Then, cognitive engineering will be used to explain the underpinnings of these interactions, thus guiding further the development of the proposed metrics. Then, we contend that together in a benchmark, these fields provide a holistic approach to creating user-centric systems where attention and judgment play critical roles in shaping front-end experiences, supported by robust back-end processes \cite{chapter4, wickens2004introduction}.

In understanding what a benchmark is and what sort of regime the field of HAI should aim at, we take the above trifecta approach (Cognitive Engineering, Human Factors, and HAI) and look at the field of AI more broadly, namely at benchmark methodologies. These methodologies are particularly relevant when considering that the evolution and evaluation of HAI systems is intertwined with AI given the increasing presence of AI algorithms across software modalities \cite{kasneci2023chatgpt, srivastava2023imitation}.

We begin then by evaluating theoretical models that have been used in cognitive engineering for describing human reasoning, namely regarding judgment and prediction because understanding these processes is essential for designing systems that effectively support human decision-making. These models provide frameworks to quantify and analyze how people perceive, interpret, and react to complex situations, enabling the development of more intuitive and responsive systems. We then analyse and explore two sources of metrics and test cases: the first, a review of automation failures in HAI \cite{Automation_failure}, and the second, a study analysing the effects of imperfect automation on decision making using a simulated task involving HAI \cite{effects_imperfect}. Then we produce a short analysis of AI's benchmark methodologies. Lastly, we propose a set of metrics that would fit in a benchmark capable of evaluating systems such as the one developed in Rovira et al., where users interact with an advisory system that supports decision making at different levels of automation \cite{effects_imperfect}.

In "The Oxford Handbook of Cognitive Engineering" by Kathleen Mosier defines judgment and prediction as processes in decision-making that assess current events and forecast future outcomes. These processes are influenced by the nature of the environment, where ambiguous probabilistic data from natural environments contrast with the deterministic data from electronic systems. The chapter outlines two strategic goals: correspondence, which seeks empirical accuracy using environmental cues, and coherence, which aims for consistency and rationality using a systematic approach to process deterministic data. There is then a contrast between intuitive processes, which are rapid and less conscious, and analytical processes that are deliberate and detailed. Mosier proposes her own metrics, albeit informally (without formulating equations), such as accuracy, reliability, and consistency to evaluate how decisions align with actual outcomes and system logic. Methods like observational studies, system simulations, and real-world applications are suggested as well to gauge the efficacy of these processes in engineered systems \cite{chapter4}.

Understanding the current definitions of judgment and prediction in the context of cognitive science, historically, may start with Egon Brunswik's Lens Model. This model details how individuals discern and decide in ambiguous settings. Central to it are three elements: distal variables (the actual attributes in the environment), proximal cues (the perceptible signals of distal variables), and judgment (the decision made based on the proximal cues) \cite{brunswik1955representative, chapter4}. As an example, consider the medical field where HAI applications involving automation can be used in diagnosis. A physician relies on perceivable signs, such as symptoms and tests to infer a patient's health condition. This model would then focus on the uncertain connection between environmental indicators and perceptions, and how perceptions inform judgments. In mapping the model directly to an HAI application, it is useful to understand how cognition is encapsulated in the process wherein external variables (e.g. patient's health condition), via a "lens" of perception (e.g. symptoms and test results), transform into internal variables that influence judgments (e.g. diagnosis) - see Section~\ref{sec:BLM}.

Judgments are not always rational, and perceptions do not always illicit accurate deductions, in fact, these are heavily altered or influenced by hundreds of fallacies and cognitive biases - a domain which has been thoroughly explored by Daniel Kahneman, Paul Slovic, and Amos Tversky. Through their research, they revealed mental shortcuts employed to streamline decision-making: heuristics which can be as misleading as they are useful. The representativeness heuristic, for example, might induce beliefs about a company's future stock performance based purely on its past. Another bias - the anchoring bias - can make an investor overly reliant on initial stock prices even in the face of new contradictory information \cite{kahneman1982judgment}.

Evaluating how cognitive heuristics and biases affect judgment and prediction is of extreme importance in the context of HAI given that these factors heavily influence human interactions with automated systems, and can lead to predictable errors in decision-making, affecting the safety and efficiency of operations \cite{kahneman1982judgment}. As such, this issue could be interpreted as an alignment problem and thus illicit measurements in the context of a benchmark  - how aligned is the HAI system to cognitive efficiency, and its inverse (see Section~\ref{sec:AS}).


We call environments where judgment and prediction is of extreme importance \textit{high-stakes stochastic environments}. Gary Klein's Naturalistic Decision Making (NDM) model has been formulated to explain decision processes in such environments, as in, for example, the stock market. Central to NDM is the Recognition-Primed Decision (RPD) model, which attempts to explain how experts make swift decisions by recognizing patterns, as well as through situational awareness, which entails understanding the current setting to anticipate future states. A common illustration of this is a firefighter at work, rapidly determining actions by identifying patterns and leveraging contextual cues. The firefighter does not assess all potential actions exhaustively but quickly identifies a feasible course based on pattern recognition, motivated by the same fallacies that Daniel Kahneman, Paul Slovic, and Amos Tversky found to hinder rational cognition \cite{kahneman1982judgment, chapter4}.

Just as in the firefighter case, most HAI applications are closely dependent to an environment, be it virtual (e.g. operative system, a virtual reality setting, the software-application environment) or in the real world (e.g. cockpit of an airplane, the burning woods, etc). Hybrid ecologies are environments where both electronic and naturalistic elements join. Effectively navigating such systems requires strategies, goals, and tactics. This makes it an opportunity for evaluation - to quantify system complexity on the back-end and the fidelity of the front-end in conveying relevant information to the operator of the HAI system. Assessing how well these models perform in practice, particularly in complex, hybrid environments where electronic and naturalistic elements intersect allows us to quantify the complexity of HAI systems both on the back-end, where algorithms and processes operate, and on the front-end, where information is conveyed to human operators \cite{chapter4}.

Hammond's Cognitive Continuum Theory (CCT) has been developed to encapsulate judgment and decision making in such hybrid environments. CCT posits that these modes are not binary, but endpoints on a cognitive spectrum. This means that depending on the task, individuals might toggle between or blend these cognitive modes together. Optimal cognitive operation is then not static; it is contingent on harmonizing the task with the most effective cognitive mode \cite{chapter4}. We propose then that CCT itself as well as other models described here can be deployed as metrics aiming at evaluating the mode or cognitive propensity of a HAI system interface (see Section~\ref{sec:CCT}, ~\ref{sec:cib}, ~\ref{sec:fe}, and ~\ref{sec:ol}).

\section{Motivation}
One of our motivations for proposing a benchmark desing relates to the current global trend of increasing automation \cite{Erdil2023ExplosiveGF}. To create a benchmark that can measure automation failure, it is important to define automation failure and to reflect on how human-automation interactions have failed in the past.

In "\textit{The failure to grasp automation failure}" , Skraaning et al. present a comprehensive analysis of prior definitions of automation failure \cite{Automation_failure}.

The authors base their definition of failure on the "Lumberjack model", as described by Wickens et al. and Onnasch et al. \cite{sebok2017implementing, onnasch2014human} (as cited by the authors of \cite{chapter4}). The model posits an abrupt decline in task performance when automation encounters failures beyond a certain, albeit vaguely defined, degree of automation threshold. This implies that high levels of automation can render the human operator inept at managing the task when the automation becomes defective. 

A more general definition, as delineated by Onnasch et al.
(as cited by the authors of \cite{chapter4}), posits that automation failures can manifest in hardware, software, or when automation is employed inappropriately \cite{onnasch2014human}. Expanding on this idea, Sebok \& Wickens suggest that a broader definition of automation failure could encompass software glitches, hardware breakdowns, and unintended applications of automation. There are also cases where automation works as per design but not in alignment with user expectations, due to an inaccurate user mental model or programming errors \cite{sebok2017implementing}.

Skraaning et al. use as an illustrative example of the consequences of such defective mental models the air crashes involving the Boeing 737 MAX aircraft. Here, a faulty sensor provided inaccurate climb angle data to an automated program within the system responsible for manoeuvring the planes' pitch (where is the plane's nose pointing?). This malfunction led the automated submodule to erroneously pull the aircraft's nose downwards to a point of no return. The challenges faced by pilots in these situations arose both from the core functionality of the automation but also because the pilots were missing a crucial parameter in their mental model of the aircraft: there was one simple button that if switched turned off the erroneous pitch controller that was led astray by the faulty sensor. This lapse in knowledge was later attributed to insufficient documentation and training surrounding the automated subsystem of the new aircraft, but ultimately the lack of back-end to front-end interaction displays (BFIDs) was a crucial factor in the accidents. Some pilots who knew of the existence of the switch managed to quickly gain control over the aircraft avoiding any catastrophes \cite{Automation_failure}.

To enable classification and analysis of such accidents, Skraaning et al. introduce a preliminary taxonomy of automation impairment mechanisms. This taxonomy comprises four categories: Elementary Automation Failures - failures at the component level where individual elements of the automation system fail to perform their designated functions; Systemic Automation Failures - failures that involve complex interactions between multiple components or systems, leading to cascading failures or unintended system behaviors; and Human-Automation Interaction Breakdowns - situations where the automation design does not adequately account for human capabilities and limitations, leading to mismatches between user expectations and system behavior. Lastly, the fourth category, "Human and Organizational Slips and Misconceptions" is suggested comprising of errors stemming from human mistakes or organizational oversights, not directly caused by technology failures but by poor human decisions or misunderstandings about the system’s functionality (see Section ~\ref{sec:TAIM} for a formulation of these metrics). In the paper, the authors discuss Indonesia AirAsia Flight 8501 as such an example. The pilot in command attempted to address a series of spurious warnings by resetting the Flight Augmentation Computer while in flight, which inadvertently disengaged the autopilot and other critical systems. This action led to a manual flight situation that the crew was not prepared to handle, culminating in a loss of control and the aircraft crashing \cite{Automation_failure}.

This event links to the other categories in a contrasting manner: while the automation systems themselves did not fail directly, the pilot’s misunderstanding and the organizational practices that did not discourage such an intervention during flight reflect gaps in the human-automation system interface design and training which, according to the authors, was instigated by human error stemming from insufficient understanding and an organizational bias towards certain alarms \cite{Automation_failure}.

We believe that the system communication errors involving the Boeing 737 MAX aircraft could have been preemptively diagnosed if such a benchmark was available given that it acknowledges the number of automated systems in the back-end and their front-end representations (see metric in Section \ref{sec:cib}, where we propose a simple formulation for this).

\section{Simulation Environments to Benchmark HAI}
In order to apply a benchmark to an HAI system, simulated environments with tests or tasks must be created in which to plug metrics that evaluate and compare different designs. Examples of such tasks can be found in the work of Cnossen, Meijman, and Rothengatter, who evaluated strategies and subjective workload of drivers using in-car navigation systems \cite{cnossen2004adaptive}. These studies provide examples of tasks that can be simulated and developed into tests for benchmarking HAI systems. We now look at the work of Parasuraman, Sheridan, and Wickens specifically, who introduced a taxonomy aimed at guiding the design of automation \cite{effects_imperfect}.

The authors begin by defining imperfect automation, basing themselves on a study by Crocoll and Coury, which demonstrated that participants with imperfect automation aids performed better when provided only with status information rather than decision support, which refers to automated systems suggesting specific decisions or actions based on data analysis  \cite{crocoll1990status}. Another study conducted by Sarter and Schroeder in a flight simulation environment confirmed these findings, showing that imperfect automation could degrade performance when it offers decision support in comparison to when it merely provides information support \cite{sarter2001supporting}.

The model proposed by Crocoll and Coury breaks down human information processing into four stages that can potentially benefit from automation support. These stages are: Information acquisition (which involves collecting raw data from various sources), Information analysis (involving the integration and synthesis of this data to form a comprehensive understanding), Decision and action selection (where decisions are made based on the analyzed information and potential actions are selected), and Action implementation (the execution of the chosen actions). They pose that the degree of automation can vary within these stages, from entirely manual to fully automated processes \cite{effects_imperfect}.

The authors describe the "first automation failure effect" as the impact of an initial automation failure on user performance. This effect arises from user complacency due to consistent exposure to flawless automation, causing poor responses to the first failure. This aligns with biases highlighted by Kahneman, Slovic, and Tversky, and is anticipated by models like the Lens model \cite{kahneman1982judgment, chapter4}. However, a study by Merlo, Wickens, and Yeh in 2003 revealed that experience with imperfect automation can adjust user trust and even improve user performance after adaptation to imperfect automation \cite{yeh2003head}.

When mapping their taxonomy to a real-world model, Crocoll and Coury propose the sensor-to-shooter loop simulation - this would be the test or task in a benchmarking context. This simulation tracks the flow of information processing where data about targets is first collected through "sensors" and visualized by the operator in a map. The operator then selects an action to be executed. The simulation reflects real-world command and control scenarios by mapping each step from detection to action. Three types of automation are provided in the simulation: information automation, which gives updates about the target's status, and decision automation, which offers recommendations on identifying the targets. Additionally, a combination of both information and decision automation is used \cite{effects_imperfect}.

More specifically, participants were given a software program tasked with identifying the most dangerous enemy target and selecting a friendly unit to engage it on a map, termed as "enemy-friendly engagement selection." The automation support was integrated into the bottom-left section of the task window and it provided the levels of support shown in Table \ref{tab:automation_levels}.

\begin{table}[H]
\tiny
\centering
\begin{tabular}{|c|p{10cm}|}
\hline
\textbf{Level of Automation} & \textbf{Description} \\ \hline
Information Automation & Displayed all possible engagement combinations, distances between targets, friendly units, and headquarters without explicitly guiding the decision process. \\ \hline
Low Decision Automation & Provided prioritized listings of all engagement combinations, categorizing it as decision automation. \\ \hline
Medium Decision Automation & Displayed only the top three engagement options. \\ \hline
High Decision Automation & Recommended the best engagement option, with some data not immediately accessible to the user. \\ \hline
\end{tabular}
\caption{Levels of Decision Automation and Their Descriptions}
\label{tab:automation_levels}
\end{table}

Four metrics were used to evaluate operators. First, decision-making accuracy was measured as the percentage of trials in which participants correctly selected the most dangerous enemy target and the appropriate friendly unit to engage it. The results showed that reliable automation did not significantly improve accuracy compared to manual performance, while unreliable automation led to a significant reduction in accuracy. Second, decision-making response time (RT) was measured as the time taken to make an engagement decision in milliseconds. Reliable automation significantly reduced decision-making RT compared to manual performance, whereas unreliable automation increased RT relative to reliable automation but not compared to manual performance. Thirdly, secondary task performance was evaluated using a communications task, where accuracy and response times were measured. The results indicated that high decision automation improved response times for the secondary task. Finally, subjective ratings of mental workload, trust, and self-confidence were collected using the NASA-Task Load Index and Likert scales. Mental workload ratings were higher under information automation, while trust ratings were higher for decision automation than for information automation. Despite variations in overall automation reliability, no significant effect was found on trust ratings \cite{effects_imperfect}. 

The participants first performed tasks without automation support, followed by tasks with varied levels of automation reliability. The study concluded that automation significantly impacts operator performance, particularly influencing the response times and accuracy of decisions in command and control tasks. Reliable automation notably improved response times for making engagement selections, demonstrating its potential to expedite decision-making when functional. However, the effectiveness of automation was compromised by its reliability. High reliability in automation led to greater costs in decision-making accuracy when automation failed, especially in scenarios involving decision automation as opposed to mere information automation (see Table \ref{tab:automation_levels}). This suggests a higher vulnerability in performance when automation is extensively involved in decision processes \cite{effects_imperfect}.

At lower levels of automation reliability, both information and decision automation negatively impacted performance, indicating that below certain reliability thresholds, automation can detrimentally affect outcomes irrespective of the automation type. Comparatively, the accuracy of engagement decisions between manual control and reliable automation showed no significant difference but declined with unreliable automation, highlighting risks of dependency on automation \cite{effects_imperfect}.

As mentioned before, the study also explored the 'first automation failure effect', recording initial poor performance when automation fails, which could potentially improve as operators adjust their reliance on automation based on experience \cite{effects_imperfect}. 

This study then shows the importance of designing automation systems that are not only accurate but also consistently reliable to effectively support critical decision-making processes \cite{effects_imperfect}. We contend that one way of pursuing such designs is through the methodology of benchmarking so as to iteratively perfect a system in a testable manner.

\section{Benchmarking in AI}
\label{sec:bench_ai}
We contend that the evolution of benchmarks in AI from Deep Learning (DL) to Deep Reinforcement Learning (DRL) and Large Language Models (LLMs) is relevant to our problem of developing a benchmark in HAI in two primary ways.

First, as AI progresses, benchmarks for HAI must be capable of evaluating AI models in the HAI system. Secondly, we do not have to reinvent the wheel - the field of AI has constructed a sound methodological theory of benchmarks on which we will base ours.

As an example of the necessity of continuous benchmarking development, Sattler et al. demonstrated the limitations of CNN-based Absolute Pose Regression models compared to handcrafted image retrieval baselines by essentially creating a new test - one that compares the two methods and shows their limitations \cite{sattler2019understanding}.

Benchmarks should then assess the explainability of system behavior, particularly in high-stakes contexts, where it is important for users to be able to rectify and make sense out of model responses. Arulkumaran et al. identified key challenges in deep reinforcement learning (DRL) models due to their black-box nature\footnote{In deep learning, a black-box model refers to an AI system whose internal workings are not interpretable or transparent to humans, making it difficult to understand how the model arrives at specific decisions or predictions \cite{sattler2019understanding}} and interpretability issues. These challenges include sample inefficiency, lack of theoretical understanding amongst users, and reproducibility issues across the field  which makes it difficult to compare models and to gauge if a model is even good at a particular task \cite{arulkumaran2017deep}.

Another recent development in AI, LLMs like the GPT model serie used in ChatGPT developed by OpenAI, have brought the attention of many researchers and their groups to the task of innovating quality metrics to assess their state of development and impact effectively. Kasneci et al. emphasize the potential of these models in education and healthcare while also highlighting ethical implications and misinformation risks. They suggest that quality metrics for LLMs should include factual accuracy, ethical considerations, and increases in user cognitive workload, giving us another example of a field where rapid technological development must be accompanied by rapid development of evaluation techniques \cite{kasneci2023chatgpt}.

Typically, in AI, the word 'benchmark' is ambiguous, referring to a test or a set of tests, their metrics, or both: the test and the associated metric. These benchmarks are used in the context of comparing different frameworks or models so as to discover which is the best performing - the state of the art - in a particular task as measured by one or many metrics \cite{lm_evaluation_harness, rajpurkar2016squad, wang2018glue, srivastava2023imitation, lintang_sutawika_2023_10256836}.

In Deep Learning benchmarks the objective has been to measure algorithmic progress in particular tasks and in terms of computation effort. In computer vision, Convolutional Neural Networks (CNNs) are evaluated using datasets like ImageNet and COCO. In natural language processing, the evaluation of Long Short Term Memory networks (LSTMs) and Recurrent Neural Networks (RNNs) consists of metrics such as accuracy, precision, recall, and F1-score for assessing classification and prediction capabilities \cite{deng2009imagenet, lin2014microsoft, sokolova2009systematic}. Deep Reinforcement Learning utilises these same metrics, but learning is evaluated differently, namely through interactive environments and by counting reward\footnote{In reinforcement learning, a reward is a signal provided to the agent, indicating the value of the actions taken, guiding it toward desired behavior in an environment \cite{mnih2013playing}.}. Environments can be, for example, the game of tic-tac-toe, the Atari 2600 games, or the game of GO \cite{mnih2013playing}. In the latter example, DeepMind's AlphaGo, an automated deep learning system, even beat the world champion in 2018 \cite{silver2016mastering}. Metrics for these environments include cumulative reward and exploration efficiency, assessing algorithms' adaptability and decision-making \cite{mnih2013playing, silver2016mastering}. 

For Large Language Models, benchmarks like SQuAD \cite{rajpurkar2016squad}, GLUE \cite{wang2018glue}, BIG-bench \cite{srivastava2023imitation}, and lm-evaluation-harness \cite{lintang_sutawika_2023_10256836} assess language comprehension, generation, and adaptability. Tasks range from multiple-choice to dialogue completion. Metrics evaluate response correctness, fluency, and context maintenance, measuring the models' response quality to language inputs \cite{rajpurkar2016squad, wang2018glue, srivastava2023imitation, lintang_sutawika_2023_10256836}.

\begin{table}[H]
\centering
\begin{tabular}{|l|l|c|}
\hline
\textbf{Metric} & \textbf{Description} & \textbf{Equation} \\ \hline
Accuracy & Ratio of correct identifications to total samples & \(\frac{TP+TN}{TP+TN+FP+FN}\) \\ \hline
Precision & Ratio of true positives to total positive results & \(\frac{TP}{TP+FP}\) \\ \hline
Recall & Ratio of true positives to actual positives & \(\frac{TP}{TP+FN}\) \\ \hline
F1-score & Harmonic mean of precision and recall & \(2 \times \frac{\text{Precision} \times \text{Recall}}{\text{Precision} + \text{Recall}}\) \\ \hline
Cumulative Reward & Sum of rewards obtained & \(\sum_{t=0}^{T} R_t\) \\ \hline

\end{tabular}
\caption[]{Commonly Used Metrics in DL, DRL, and LLMs.\protect\footnotemark}
\end{table}
\footnotetext{Where TP = True Positives, TN = True Negatives, FP = False Positives, FN = False Negatives, and \(R_t\) = reward at time \(t\).}






\section{$\Delta$-EVAL}\label{sec:HAIBENCH}

A benchmark in HAI must have as one of its goals guiding design towards the most cognitively aligned or optimised system so as to minimize accidents or unwanted consequences and maximise the desired outcomes resulting from the human-machine interaction. There are at least two obvious dimensions to this problem: the human operator, and the machine \cite{chapter4}. We incorporate both aspects of the problem in our abstract development of the HAI benchmark we are calling $\Delta$-EVAL. In Table \ref{tab:hypothetical_bench} we apply the general principles we extrapolated from the AI benchmarks reviewed in Section \ref{sec:bench_ai} to guide the creation of the metrics proposed under Section \ref{sec:systemmetrics}.



Each metric, when applied to a task, should aim at facilitating the evaluation of one of these aspects, or both, depending on the context and application. We do not explore explicitly the evaluation of specific automation algorithms that would fall under the scope of AI algorithms such as DL, DRL, LLMs, or more broadly speaking, Machine Learning (ML), since these already have an existing standard of evaluation form \cite{lm_evaluation_harness, rajpurkar2016squad, wang2018glue, srivastava2023imitation, lintang_sutawika_2023_10256836}. The different aspects of the problem, namely the operator and the machine or software system, can be broadly categorised given the metrics that apply to them and we explicitly define them as isolated modules and explore them further in Section~\ref{sec:framework_key}. 

Furthermore, we abstract some principles from the AI benchmark methodology in Table \ref{tab:hypothetical_bench} in order to apply them to our HAI benchmark. This distilled framework consists of goals, metrics, tasks, and optionally, relevant datasets. In the sec we have developed many examples of metrics and use cases based on this. We also provide one example that should illustrate how to directly apply the framework to HAI in Section~\ref{sec:example_framework_application}.

\begin{table}[H]
\tiny
\centering
\begin{tabular}{|>{\raggedright\arraybackslash}m{3cm}|>{\raggedright\arraybackslash}m{9cm}|}
\hline
\textbf{Benchmark Component} & \textbf{Description} \\ \hline
Metrics & Choose suitable metrics, both quantitative (e.g., accuracy, response time) and qualitative (e.g., user stress or ease of use). Avoid Goodhart's law, where a metric that becomes a target ceases to be a good metric. \\ \hline
Tasks & Design tests for evaluating the HAI system against chosen metrics, covering algorithmic challenges, user interactions, system stability, and module integration, embodying the set goals (e.g., piloting a plane safely). \\ \hline
(Optional) Relevant Datasets & Select datasets that mirror real-world scenarios relevant to the HAI system's use. Automated generators may be used, but testing simulated HAI with human operators and collecting that data for later use is often more appropriate. \\ \hline
Iterative Refinement & Continuously refine the benchmark through iterative testing, adapting measurement methods, creating new tests, generating and collecting datasets, and making comparisons across system designs and rules. \\ \hline
Standardization and Comparability & Produce a standardized design in benchmarks to ensure broad applicability and comparability across different HAI systems, using recognized metrics, common datasets, and standardized testing procedures or tasks. \\ \hline
\end{tabular}
\caption{Hypothesized Benchmark Design Components for HAI Systems}
\label{tab:hypothetical_bench}
\end{table}

\section{Methodology}

\subsection{Designing $\Delta$-EVAL}
\label{sec:systemmetrics}
Metrics can be measured in an HAI system just like the one developed in the study by Rovira et al., where decision-making accuracy, response time, secondary task performance, and subjective ratings of workload, trust, and self-confidence were measured to evaluate the effects of different types (information vs. decision automation) and levels (60\% vs. 80\% reliability) of automation on human performance in a simulated command and control task \cite{effects_imperfect}. An exercise in translating existing Cognitive Engineering models into possible metrics can be found in Section~\ref{sec:TCE}, where models such as Policy Capturing (PC) \cite{Slovic1969AnalyzingTE}, Signal Detection Theory (SDT) \cite{Green1966SignalDT}, Naturalistic Decision Making (NDM) \cite{Klein1993ARD}, Coherence Model (CM) \cite{Hogarth1992OrderEI}, Cognitive Continuum Theory (CCT) \cite{Hammond1999JudgmentsUS}, and Brunswik's Lens Model (BLM) \cite{brunswik1955representative} are translated into metrics deployable in a benchmark. Each of these models is provided with specific variables, mathematical formulations, and application examples, such as aircraft piloting, air traffic control, automated driving systems, robotic surgery, and spacecraft monitoring systems, demonstrating their practical application in evaluating HAI systems. These, and simpler hypothetical ones (also in the Section), serve the purpose of quantifying and guiding the design of HAI systems and principals in Human Factors or Cognitive Engineering as per \cite{Automation_failure, chapter4, effects_imperfect}. We see this as a start in what can become a very lengthy exercise in creativity when generating such metrics (EleutherAI's LLM benchmark as of the time of writing has 47 active tests while BIG-bench lists more than 200 tasks such as tic-tac-toe, chess, completion questions in French, multiple-choice questions about biology and many more, all of which have their own ways of measuring outcomes \cite{lm_evaluation_harness, srivastava2023imitation}). We hope to have grounded the proposed metrics sufficiently on well-established theoretical models prominent in Cognitive Science so as to build off of its widely accepted axioms.

\subsection{Translating Cognitive Engineering models into Evals}
\label{sec:TCE}

\subsubsection{Policy Capturing (PC) in Aircraft Piloting}
\label{sec:PC}
Policy Capturing (PC) methodology models strategies used by pilots in making judgments and predictions based on cockpit cues.\\

\textbf{Variables}:
\begin{itemize}
    \item $C_{j}$: Cue $j$ provided to the pilot (e.g., instrument readings, weather information).
    \item $W_{j}$: Weight assigned to cue $j$ indicating its importance in decision-making.
    \item $D$: Decision made by the pilot (e.g., change course, altitude adjustments).
\end{itemize}

\begin{equation}
D = \sum_{j} W_{j} \times C_{j}
\end{equation}

\textbf{Example in Piloting}\\
This linear-additive model represents the pilot's judgment process, where each cue's contribution to the decision is weighted.
In a simulated flight scenario, cues ($C_{j}$) such as altitude, speed, and external weather data are presented to the pilot. The pilot's decisions ($D$) are recorded, and the weights ($W_{j}$) are derived through regression analysis to understand how different cues influenced the decision.

\subsubsection{Signal Detection Theory (SDT) Eval}
\label{sec:SDT}
\begin{itemize}
    \item $H$: Number of Hits
    \item $CR$: Number of Correct Rejections
    \item $FA$: Number of False Alarms
    \item $M$: Number of Misses
    \item $N_{\text{total}}$: Total Number of Judgements
    \item $d'$: Sensitivity
    \item $c$: Response Bias
    \item $Z(H)$: This refers to the Z-score of the hit rate. The hit rate is the proportion of correctly identified positive cases (e.g., correctly identifying a signal when it is present). The Z-score transformation of the hit rate normalizes it relative to the standard normal distribution.
    \item $Z(FA)$: This is the Z-score of the false alarm rate. The false alarm rate is the proportion of negative cases incorrectly identified as positive (e.g., incorrectly identifying a signal when it is absent). The Z-score for the false alarm rate similarly normalizes this value.
\end{itemize}
\begin{equation}
\text{SDT Score} = \frac{H + CR}{N_{\text{total}}}, \quad d' = Z(H) - Z(FA), \quad c = -\frac{1}{2}(Z(H) + Z(FA))
\end{equation}

\textbf{Example in Air Traffic Control}\\
In air traffic control, the SDT metric can be applied to evaluate the accuracy of recognizing aircraft on radar screens. Hits ($H$) could be correctly identified aircraft, while Correct Rejections ($CR$) are accurately dismissing non-aircraft signals. False Alarms ($FA$) occur when controllers mistakenly identify a non-aircraft object as an aircraft, and Misses ($M$) are when actual aircraft are overlooked. The total number of judgments ($N_{\text{total}}$) would be the sum of all these instances. Sensitivity ($d'$) and Response Bias ($c$) are calculated based on the rates of Hits and False Alarms, evaluating the controllers' detection capabilities and biases towards safe or risky decisions.

\subsubsection{Naturalistic Decision Making (NDM) Eval}
\label{sec:NDM}
\begin{itemize}
    \item $D_{\text{efficient}}$: Number of Efficient Decisions
    \item $D_{\text{total}}$: Total Number of Decisions
    \item $T_{\text{decision}}$: Time taken for Decision-Making
\end{itemize}
\begin{equation}
\text{NDM Score} = \frac{D_{\text{efficient}}}{D_{\text{total}}}, \quad \text{Speed of Decision-Making} = \frac{1}{T_{\text{decision}}}
\end{equation}
\textbf{Example in Piloting Aircraft}\\
For pilots, the NDM metric can measure decision-making efficiency during critical flight situations. Efficient Decisions ($D_{\text{efficient}}$) might involve correct navigational choices or emergency responses, while the Total Decisions ($D_{\text{total}}$) would include all decisions made during a flight. The Time taken for Decision-Making ($T_{\text{decision}}$) can be observed during simulated emergency scenarios to calculate the Speed of Decision-Making, assessing how quickly pilots can make accurate decisions under pressure.

\subsubsection{Coherence Model Eval}
\label{sec:CM}
\begin{itemize}
    \item $J_{\text{coherent}}$: Number of Coherent Judgements
    \item $J_{\text{total}}$: Total Number of Judgements
    \item $B_{\text{assessment}}$: Bias Assessment Score
\end{itemize}
\begin{equation}
\text{Coherence Score} = \frac{J_{\text{coherent}}}{J_{\text{total}}}, \quad B_{\text{assessment}} = \text{Function of Bias Detection and Correction}
\end{equation}
\textbf{Example in Automated Driving Systems}\\
In an automated driving system, an operator (such as a safety supervisor or a remote controller) monitors and intervenes in the system’s decisions. Coherent Judgements ($J_{\text{coherent}}$) are instances where the operator accurately assesses and possibly overrides the system’s decisions, ensuring logical and safe navigation. The Total Judgements ($J_{\text{total}}$) include all interactions the operator has with the system’s decisions. The Bias Assessment ($B_{\text{assessment}}$) involves evaluating how effectively the operator identifies and corrects potential biases in the system’s decision-making, such as over-reliance on certain sensors or data inputs. This metric assesses the operator’s effectiveness in maintaining coherent and safe decision-making in collaboration with the automated system.

\subsection{Cognitive Continuum Theory (CCT) Eval for Robotic Surgery}
\label{sec:CCT}
\textbf{Variables:}
\begin{itemize}
    \item $T_{t}$: Time taken for test $t$
    \item $N$: Number of Tests
    \item $S_{t}$: Spectrum Score for test $t$ (ranging from -1 for intuitive to 1 for analytical)
\end{itemize}
\textbf{CCT}:
\begin{align}
I_{t} &= \frac{1}{T_{t}} \quad \text{(Intuitive Decision Score, inversely proportional to time)} \\
A_{t} &= \frac{1}{I_{t}} \quad \text{(Analytical Decision Score, inversely proportional to intuitive score)} \\
S_{t} &= A_{t} - I_{t} \quad \text{(Spectrum Score, the balance between intuitive and analytical)} \\
\text{CCT Score} &= \frac{\sum_{t=1}^{N} S_{t}}{N} \\
\end{align}

\textbf{Example in Robotic Surgery}\\
In robotic surgery, each surgery or simulation is a test. Intuitive Decision Score ($I_{t}$) is calculated as the reciprocal of the time taken for the surgeon to make a decision, where quicker responses indicate higher intuitiveness. Analytical Decision Score ($A_{t}$) is the reciprocal of the Intuitive Decision Score, representing analytical decision-making as the inverse of intuitiveness. The Spectrum Score ($S_{t}$) for each test reflects the balance between intuitive and analytical decisions, with positive values indicating a more analytical approach and negative values indicating a more intuitive approach. The overall CCT Score is the average of Spectrum Scores across multiple tests, providing information about how surgeons blend these cognitive modes during surgeries.

\subsection{Brunswik's Lens Model Eval}
\label{sec:BLM}
\begin{itemize}
    \item $J_{\text{accurate}}$: Number of Accurate Judgements based on Cues
    \item $J_{\text{total}}$: Total Number of Judgements
    \item $E_{\text{validity}}$: Ecological Validity Assessment
\end{itemize}
\begin{equation}
\text{Lens Score} = \frac{J_{\text{accurate}}}{J_{\text{total}}}, \quad E_{\text{validity}} = \text{Assessment of Cue Interpretation Accuracy}
\end{equation}
\textbf{Example in Spacecraft Monitoring Systems}\\
In spacecraft monitoring, the Lens Model can be applied to evaluate how operators interpret telemetry data. Accurate Judgements ($J_{\text{accurate}}$) are cases where operators correctly interpret data to identify the state of the spacecraft. Total Judgements ($J_{\text{total}}$) include all instances of data interpretation. Ecological Validity Assessment ($E_{\text{validity}}$) examines how well the interpreted data reflects the actual state of the spacecraft, ensuring operators are making decisions based on accurate and relevant information.


\subsection{Cognitive Alignment}
\label{sec:AS}
\begin{align}
\text{Let:} \\
& h_i &: \text{Heuristic } i \\
& T_{h_i} &: \text{Test or set of tests for heuristic } h_i \\
& TP_{h_i}(T_{h_i}) &: \text{True Positives for heuristic } h_i \text{ using test } T_{h_i} \\
& TN_{h_i}(T_{h_i}) &: \text{True Negatives for heuristic } h_i \text{ using test } T_{h_i} \\
& FP_{h_i}(T_{h_i}) &: \text{False Positives for heuristic } h_i \text{ using test } T_{h_i} \\
& FN_{h_i}(T_{h_i}) &: \text{False Negatives for heuristic } h_i \text{ using test } T_{h_i} \\
& H &: \text{Total number of heuristics considered} \\
\end{align}

Define the Heuristic Triggering Score (HTS) for each heuristic:

\begin{align}
\text{Let:} \\
& M_{TP_{h_i}} = \text{Mean}(TP_{h_i}(T_{h_i})) \\
& M_{TN_{h_i}} = \text{Mean}(TN_{h_i}(T_{h_i})) \\
& M_{FP_{h_i}} = \text{Mean}(FP_{h_i}(T_{h_i})) \\
& M_{FN_{h_i}} = \text{Mean}(FN_{h_i}(T_{h_i})) \\
\end{align}
Then, the Heuristic Triggering Score (HTS) for heuristic $h_i$ \text{ is:}

\begin{equation}
\text{HTS}_{h_i} = \frac{M_{TP_{h_i}} + M_{TN_{h_i}} - M_{FP_{h_i}} - M_{FN_{h_i}}}{M_{TP_{h_i}} + M_{TN_{h_i}} + M_{FP_{h_i}} + M_{FN_{h_i}}}
\end{equation}

\text{The overall system's Normalized Heuristic Triggering Index (NHTI) is calculated as:}

\begin{equation}
AS = \frac{1}{H} \sum_{i=1}^{H} HTS_{h_i}
\end{equation}

\textbf{Aircraft Piloting Example} \\
In this scenario pilots are responding to various cockpit alarms and indicators during a flight.
For each heuristic $h_i$, true positives ($TP_{h_i}$) could be correct responses to alarms, true negatives ($TN_{h_i}$) are ignoring irrelevant or false alarms, false positives ($FP_{h_i}$) are overreacting to non-critical alarms, and false negatives ($FN_{h_i}$) are missing critical alarms. The Heuristic Triggering Score ($\text{HTS}_{h_i}$) would be calculated based on these values for each heuristic encountered during the flight, leading to an overall Cognitive Alignment score.

\subsection{Taxonomy of automation impairment mechanisms}
\label{sec:TAIM}
Based on the work by Skraaning et al. \cite{Automation_failure}.
\subsection{Elementary Automation Failures (EAF)}
\begin{itemize}
\item Let $C = {c_1, c_2, \ldots, c_n}$ be the set of all components in the automation system.
\item For each component $c_i$, define:
\begin{itemize}
\item $f_{c_i}$: The designated function of component $c_i$
\item $p_{c_i}$: The observed performance of component $c_i$
\item $w_{c_i}$: The criticality weight of component $c_i$, based on its potential impact or consequences
\item $\epsilon_{c_i} = \begin{cases}
w_{c_i}, & \text{if } p_{c_i} \neq f_{c_i} \
0, \text{otherwise}
\end{cases}$
\end{itemize}
\end{itemize}
Then, the Weighted Elementary Automation Failure (WEAF) score is defined as:
\begin{equation}
\text{WEAF} = \frac{1}{\sum_{i=1}^{n} w_{c_i}} \sum_{i=1}^{n} \epsilon_{c_i}
\end{equation}
The WEAF score represents the proportion of components in the system that failed to perform their designated functions correctly, weighted by their criticality. Components with higher criticality weights contribute more to the overall score.
The Weighted Elementary Automation Failure (WEAF) score can be normalized by dividing it by the maximum possible score, which occurs when all components fail:
\begin{equation}
\text{WEAF}{\text{normalized}} = \frac{\text{WEAF}}{\sum{i=1}^{n} w_{c_i}}
\end{equation}
This normalized WEAF score will range from 0 (no component failures) to 1 (all components failed, weighted by their criticality).

\subsection{Systemic Automation Failures (SAF)}
\begin{itemize}
\item Let $\mathcal{S} = {S_1, S_2, \ldots, S_m}$ be the set of all subsystems or interacting components in the automation system.
\item For each subsystem $S_j$, define:
\begin{itemize}
\item $g_{S_j}$: The expected behavior of subsystem $S_j$
\item $q_{S_j}$: The observed behavior of subsystem $S_j$
\item $s_{S_j}$: The severity level of a failure in subsystem $S_j$, based on its potential consequences
\item $\delta_{S_j} = \begin{cases}
s_{S_j}, & \text{if } q_{S_j} \neq g_{S_j} \
0, \text{otherwise}
\end{cases}$
\end{itemize}
\end{itemize}
Then, the Weighted Systemic Automation Failure (WSAF) score is defined as:
\begin{equation}
\text{WSAF} = \frac{1}{\sum_{j=1}^{m} s_{S_j}} \sum_{j=1}^{m} \delta_{S_j}
\end{equation}
The WSAF score represents the proportion of subsystems or interacting components in the system that exhibited unintended or cascading failures due to complex interactions, weighted by the severity level of each subsystem failure.
Similarly, the Weighted Systemic Automation Failure (WSAF) score can be normalized by dividing it by the maximum possible score, which occurs when all subsystems fail at their highest severity level:
\begin{equation}
\text{WSAF}{\text{normalized}} = \frac{\text{WSAF}}{\sum{j=1}^{m} \max(s_{S_j})}
\end{equation}
This normalized WSAF score will range from 0 (no subsystem failures) to 1 (all subsystems failed at the highest severity level).

\subsection{Human-Automation Interaction Breakdowns (HAIB)}
\begin{itemize}
\item Let $\mathcal{T} = {T_1, T_2, \ldots, T_k}$ be the set of all tasks or interactions between the human operator and the automation system.
\item For each task $T_l$, define:
\begin{itemize}
\item $h_{T_l}$: The expected human performance for task $T_l$
\item $r_{T_l}$: The observed human performance for task $T_l$
\item $c_{T_l}$: The criticality weight of task $T_l$, based on its potential consequences
\item $\gamma_{T_l} = \begin{cases}
c_{T_l}, & \text{if } r_{T_l} \neq h_{T_l} \
0, \text{otherwise}
\end{cases}$
\end{itemize}
\end{itemize}
Then, the Weighted Human-Automation Interaction Breakdown (WHAIB) score is defined as:
\begin{equation}
\text{WHAIB} = \frac{1}{\sum_{l=1}^{k} c_{T_l}} \sum_{l=1}^{k} \gamma_{T_l}
\end{equation}
The WHAIB score represents the proportion of tasks or interactions where the human operator's performance deviated from the expected performance due to mismatches between user expectations and system behavior, weighted by the criticality of each task or interaction.
The Weighted Human-Automation Interaction Breakdown (WHAIB) score can be normalized by dividing it by the maximum possible score, which occurs when all tasks or interactions result in breakdowns at the highest criticality level:
\begin{equation}
\text{WHAIB}{\text{normalized}} = \frac{\text{WHAIB}}{\sum{l=1}^{k} \max(c_{T_l})}
\end{equation}
This normalized WHAIB score will range from 0 (no interaction breakdowns) to 1 (all interactions resulted in breakdowns at the highest criticality level).

\section{Component Interaction Balance (CIB)}
\label{sec:cib}
Measures the ratio of front-end components to back-end interactions.\\

\begin{equation} CIB = \frac{\text{Number of front-end components}}{\text{Number of back-end interactions}} \end{equation}
\textbf{Ideal Value}: Close to 1, indicating a balanced system.

\textbf{Aircraft Piloting Example:}
Here, pilots are interacting with cockpit instruments (front-end components) and the aircraft's response systems (back-end interactions).
$CIB$ is calculated as the ratio of the number of instrument interactions (front-end components) to the number of automatic responses by the aircraft (back-end interactions). An ideal value close to 1 indicates a well-balanced interaction between pilot input and aircraft response.

\subsection{Overload Penalty (OP)}
\label{sec:op}
Quantifies the excess of back-end interactions compared to front-end components. \\

\begin{equation}
OP = \max(0, \text{Number of back-end interactions} - \text{Number of front-end components})
\end{equation}

\textbf{Aircraft Piloting Example:}
Let's take now pilots that are managing multiple tasks and system responses during critical flight phases.
$OP$ quantifies the excess of system responses (back-end interactions) compared to the tasks managed by the pilot (front-end components). This metric helps identify if pilots are overloaded by excessive automated responses during flight operations.

\subsection{Attention Span Efficiency (ASE)}
\label{sec:ase}
Evaluates how front-end components are chunked, considering optimal human working memory capacity. Based on chunking theory, this metric assesses if the system design facilitates efficient cognitive processing. \\

\begin{equation}
ASE = \frac{\text{Number of effectively chunked components}}{\text{Total number of front-end components}}
\end{equation}

\textbf{Conditions for Chunking} based on \cite{cowan2001magical}:
\begin{enumerate}
\item Information overload limits chunks to individual stimulus items.
\item Steps are taken to block the recoding of stimulus items into larger chunks.
\item Performance discontinuities caused by the capacity limit.
\item Indirect effects of the capacity limit.
\end{enumerate}

ASE close to 1 indicates effective chunking and cognitive processing, while values significantly less than 1 indicate potential cognitive overload or underutilization.

\textbf{Aircraft Piloting Example:}
In this scenario, pilots are monitoring various cockpit displays and indicators.
$ASE$ evaluates if the number of monitored components (like displays and indicators) is effectively chunked within the pilot’s working memory capacity. This score indicates if cockpit design facilitates efficient cognitive processing.

\subsection{Wasted Attention Resource (WAR)}
\label{sec:war}
Measures underutilization of human's attention if less than 5 front-end components. \\

\begin{equation} WAR = \max(0, 5 - \text{Number of front-end components}) \end{equation}

\textbf{Aircraft Piloting Example:}
\begin{enumerate}
\item Scenario: Assessing the cockpit design for optimal information presentation.
\item Measurement: $WAR$ measures underutilization of pilot attention if the number of front-end components (like displays and indicators) is less than 5, while $NI$ measures surplus components beyond the average processing capability of 9.
\end{enumerate}

\subsection{Noise Index (NI)}
\label{sec:ni}
Surplus front-end components beyond human's average processing capability. \\

\begin{equation} NI = \max(0, \text{Number of front-end components} - 9) \end{equation}

\textbf{Aircraft Piloting Example:}
Take a high-stress situation where pilots are managing cockpit information, like complex weather conditions or during critical flight phases.
In this scenario, the cockpit is equipped with various information displays, control systems, and indicators, each representing a front-end component. The Noise Index ($NI$) is calculated to evaluate if the number of these components exceeds the average processing capability of the human mind, which is typically around 9 items. This metric is attempting to determine whether the cockpit layout leads to information overload. The $NI$ is calculated using the formula.

\subsection{Interaction Redundancy (IR)}
\label{sec:ir}
Evaluates duplicated back-end interactions for a single front-end component. \\

\begin{equation} IR = \text{Total back-end interactions} - \text{Unique back-end interactions} \end{equation}

\textbf{Aircraft Piloting Example:}
When pilots are interacting with redundant systems in the cockpit, $IR$ evaluates the number of duplicated back-end interactions for a single front-end component, helping identify unnecessary system redundancies.

\subsection{Cognitive Strain Index (CSI)}
\label{sec:csi}
The CSI analyzes the cognitive load the HAI imposes on the user by evaluating task completion time and error rate.

\begin{equation}
CSI = \alpha \times \frac{TCT}{BT} + \beta \times ER
\end{equation}

Where:
\begin{itemize}
    \item $TCT$ is the Task Completion Time.
    \item $BT$ is the Eval Time, representing the ideal or average completion time for the task.
    \item $ER$ is the Error Rate, defined as the ratio of tasks with errors to the total number of tasks.
    \item $\alpha$ and $\beta$ are weighting factors. They can be adjusted based on the specific importance of time and error in the system under evaluation.
\end{itemize}

The intuitiveness of front-end components can also play a crucial role in user performance, we designed Component Clarity Score (CCS) to evaluate this aspect.

\textbf{Aircraft Piloting Example:}
Take now pilots that are performing navigational tasks under time pressure.
$CSI$ is calculated using the Task Completion Time ($TCT$) against a Eval Time ($BT$), and the Error Rate ($ER$). This could potentially reveal the cognitive load experienced by pilots during specific tasks.\footnote{\textbf{Component Clarity Score (CCS)}: 

The CCS evaluates the intuitiveness of front-end components based on users' response times and accuracy.

\begin{equation}
CCS_1 = \gamma \times \frac{ART}{ERT} + \delta \times AR
\end{equation}

Where:
\begin{itemize}
    \item $ART$ is the Average Response Time of users to a front-end component.
    \item $ERT$ is the Eval Response Time, which represents the ideal or average response time for a similar component in familiar systems.
    \item $AR$ is the Accuracy Rate, representing the ratio of correct responses to total responses.
    \item $\gamma$ and $\delta$ are weighting factors. They can be adjusted based on the specific importance of response time and accuracy for the front-end component under evaluation.
\end{itemize}

\begin{equation}
CCS_2 = \frac{\sum_{i=1}^{n} s_i}{n}
\end{equation}

Where \( s_i \) represents the clarity score (usually on a scale of 1-5 or 1-10) given by the i-th user out of \( n \) users. \\
\textbf{Ideal Value}: Higher values are preferred, indicating more intuitive components.
}

\subsection{Feedback Efficiency (FE)}
\label{sec:fe}
How many back-end interactions provide direct feedback on the front-end. \\
\begin{equation} FE = \frac{\text{Number of back-end feedback interactions}}{\text{Total number of back-end interactions}} \end{equation}

\textbf{Aircraft Piloting Example:}
\begin{enumerate}
\item Scenario: Evaluating the effectiveness of the aircraft's feedback systems, such as warning lights or autopilot adjustments.
\item Measurement: In this scenario, $FE$ is calculated by dividing the number of back-end interactions that result in direct feedback to the pilot (like audible alerts, visual signals on the dashboard, or tactile feedback) by the total number of back-end interactions (which includes all automated system responses). This metric assesses how effectively the aircraft's systems communicate critical information back to the pilot. An $FE$ score close to 1 would indicate that most back-end interactions are effectively communicated to the pilot, enhancing situational awareness and decision-making.
\end{enumerate}

\subsection{Operational Latency (OL)}
\label{sec:ol}
Evaluates the time difference between a user's action in the front-end and the system's response in the back-end. \\
\begin{equation} OL = t_{\text{response}} - t_{\text{action}} \end{equation}
Where \( t_{\text{action}} \) is the timestamp when the user took an action and \( t_{\text{response}} \) is the timestamp of the system's corresponding response. \\
\textbf{Ideal Value}: As close to 0 as possible, indicating instantaneous response.

\textbf{Aircraft Piloting Example:}
\begin{enumerate}
\item Scenario: Measuring the response time of the aircraft’s systems to pilot inputs.
\item Measurement: $OL$ is the time difference between the pilot's action ($t_{\text{action}}$) and the aircraft's system response ($t_{\text{response}}$), indicating the system's responsiveness to pilot inputs.
\end{enumerate}

\subsection{Critical Risk Index (CRI)}
\label{sec:cri}
Risk of overlooked interactions leading to catastrophic outcomes. \\
\begin{equation} CRI = 2^{\max(0, \text{Number of critical overlooked interactions} - 9)} \end{equation}

\textbf{Aircraft Piloting Example:}
In this example the metric is used in assessing the risk of critical system interactions being overlooked by pilots.
$CRI$ is calculated based on the number of critical interactions overlooked by pilots, emphasizing the risk of catastrophic outcomes due to oversight.

\subsection{Framework modules}
\label{sec:framework_key}
\textbf{Definitions:}
\label{sec:key}
\begin{enumerate}
\item $A_i$: Level of automation in stage $i$ (ranging from manual to fully automated).
\item $H_{ij}$: Human performance metric in stage $i$ under automation level $j$.
\item $S_{ij}$: System performance metric in stage $i$ under automation level $j$.
\item $F_{\text{type}}$: Types of automation failures, including software, hardware, inappropriate use, and mismatch.
\item $L_{\text{threshold}}$: Threshold level for the Lumberjack model, indicating a steep decline in performance at high levels of automation.
\item $C_{ij}$: Cognitive load on the human operator at stage $i$ under automation level $j$.
\item $H_{\text{error}}$: Human error factor at stage $i$ under automation level $j$.
\item $BFID$: Number of back-end to front-end interaction displays.
\item $\alpha_1, \beta_1, \delta_1$: Coefficients representing interaction effects between variables in the model.
\end{enumerate}

\subsubsection{Human Performance Module}
\label{sec:human_performance_module}
Integration of the Lumberjack model could be done to monitor the cognitive effect at high automation levels as per \cite{effects_imperfect}: $H_{ij} = \alpha_1 A_i - \delta_1 L_{\text{threshold}}$.
Human error and cognitive load can be measured by: $H_{ij} = \alpha_1 A_i + \beta_1 C_{ij} - H_{\text{error}}$.

\subsubsection{System Performance Module}
\label{sec:system_performance_module}
Here we account for different types of automation failures as per \cite{Automation_failure}: $S_{ij} = \alpha_2 F_{ij} - \sum F_{\text{type}}$, and incorporate BFID (see Section ~\ref{sec:framework_key}) for quality of information flow: $S_{ij} = \alpha_2 F_{ij} + BFID - \sum F_{\text{type}}$.

\textbf{Example application of $\Delta$-EVAL on the SDT Eval}\\
\label{sec:example_framework_application}
\begin{enumerate}
    \item \textbf{Metrics or Evaluations}: 
    \begin{itemize}
        \item Quantitative Metrics:
        \begin{itemize}
            \item Hit Rate: The proportion of actual signals correctly identified.
            \item False Alarm Rate: The proportion of non-signals incorrectly identified as signals.
            \item Correct Rejection Rate: The proportion of non-signals correctly identified.
            \item Miss Rate: The proportion of actual signals missed.
        \end{itemize}
        \item Qualitative Metrics:
        \begin{itemize}
            \item Operator Stress Level: Assessed through subjective scales (e.g., Likert scale).
            \item Decision Confidence: Measured by self-reporting scales post-task.
        \end{itemize}
    \end{itemize}

    \item \textbf{Tasks}: 
    Designing simulation tasks that replicate real-world air traffic control scenarios:
    \begin{itemize}
        \item Scenario-Based Simulations: Tasks involving monitoring a radar screen to identify aircraft among noise.
        \item Stress Testing: Introducing varied levels of signal complexity and environmental factors to assess performance under pressure.
        \item Time-Based Assessment: Evaluating the speed of decision-making in critical situations.
    \end{itemize}

    \item \textbf{Relevant Datasets (Optional)}: 
    Utilizing datasets that closely mirror real-world air traffic control environments:
    \begin{itemize}
        \item Historical Air Traffic Data: Including various signal types, densities, and real incident reports.
        \item Simulated Scenarios: Generated to represent a wide range of possible conditions, from normal to emergency situations.
    \end{itemize}

    \item \textbf{Iterative Refinement}: 
    Enhancing the evaluation process through continuous analysis:
    \begin{itemize}
        \item Performance Analysis: Regular assessment of controller performance metrics.
        \item Task Modification: Adjusting the complexity and nature of tasks based on data-driven insights.
        \item Feedback Incorporation: Utilizing operator feedback to refine tasks and measurement methods.
    \end{itemize}

    \item \textbf{Standardization and Comparability}: 
    Establishing uniform standards for tests and metrics to ensure consistency:
    \begin{itemize}
        \item Test Standardization: Developing a set of standardized tasks with varying levels of complexity.
        \item Metric Uniformity: Applying the same set of quantitative and qualitative metrics across all tests.
        \item Cross-Scenario Analysis: Enabling comparability of controller abilities across different air traffic scenarios.
    \end{itemize}
\end{enumerate}

\subsection{Causal Graphs}
In our proposed evaluation framework, we suggest that incorporating causal graphs could provide a promising way to link diverse cognitive engineering metrics with specific design interventions in HAI systems. By mapping out relationships between intervention variables (such as adjustments in front-end components), cognitive outcomes (like decision accuracy or cognitive load), and potential confounders (e.g., user expertise), causal graphs may offer a structured means to visualize and hypothesize about the underlying interactions within the system.

The idea is to tentatively explore how methods from causal inference, such as Pearl’s do-calculus and the back-door adjustment (see Appendix~\ref{app:causal}), might be applied to disentangle the direct and indirect effects of design changes in HAI. For example, by representing the system through a Directed Acyclic Graph (DAG), one could hypothesize whether observed improvements in cognitive performance are directly attributable to a design intervention or indirectly mediated through other factors, such as enhanced feedback efficiency or better attention management.

To apply DAGs in the framework in practice:
\begin{enumerate}
    \item Identify the intervention variable $X$ (e.g., a modification in the number of front-end components) and the cognitive outcome $Y$ (e.g., cognitive load or decision accuracy).
    \item Determine the set of confounders $Z$ (e.g., user expertise, baseline performance).
    \item Estimate the conditional distribution $P(Y \mid X, Z)$ from observational data.
    \item Use Eq.~\eqref{eq:backdoor} to compute $P(Y \mid do(X=x))$, and thereby evaluate the Average Treatment Effect via Eq.~\eqref{eq:ate}.
    \item If a mediator $M$ is present, decompose the effect of $X$ on $Y$ into direct and indirect components as described above.
\end{enumerate}

A more thorough introduction and presentation of the technique can be found in Appendix ~\ref{app:causal}.

\section{Results}
$\Delta$-EVAL follows a common pattern of quantifying different aspects of human-automation interaction systems by defining relevant variables, counting or measuring specific events or quantities, and combining them through summations, ratios, or other mathematical operations.

A fundamental step of $\Delta$-EVAL is to define the key components and interactions within the HAI system under evaluation, establishing the elements that will be measured and quantified. For each component, task, or interaction, various events or quantities are defined and counted or measured. Then, the counted or measured events are combined using summations and ratios to derive the final metric scores. In some cases, the derived metric scores are normalized or combined into overall indices to facilitate interpretation and comparison across different systems or scenarios.

Any particular implementation of $\Delta$-EVAL must be able to count and aggregate events across the HAI system. This is necessary for quantifying occurrences such as component failures, subsystem failures, interaction breakdowns, hits, misses, false alarms, efficient decisions, and coherent judgments. Using summations effectively aggregates these events across multiple components, subsystems, or interactions. Proportions and rates capture the proportion or rate of certain events relative to the total number of opportunities or instances. Ratios and proportions are interpretable methods for quantifying the frequency or prevalence of specific events within a context or sample size. And finally, $\Delta$-EVAL must apply weighting factors when studying the combined effects of multiple quantities or co-factors while interpreting different importance or priority levels. $\Delta$-EVAL is effectively an operationalization of the theories that underlie HAI, Human Factors, and Cognitive Engineering, allowing for the combination of diverse and holistic multivariate analysis with varying degrees of freedom enabled by the use of coefficients or normalization. In Tables \ref{tab:CIB_tab}, \ref{tab:OP_tab}, \ref{tab:ASE_tab}, \ref{tab:CSI_tab}, \ref{tab:NI_tab}, \ref{tab:IR_tab}, and \ref{tab:CRI_tab}, we introduce a sample of possible metrics to include in $\Delta$-EVAL.

\begin{table}[H]
\tiny
\centering
\begin{tabular}{|>{\raggedright\arraybackslash}m{0.5cm}|>{\raggedright\arraybackslash}m{2cm}|>{\centering\arraybackslash}m{3cm}|>{\raggedright\arraybackslash}m{3cm}|>{\raggedright\arraybackslash}m{3cm}|}
\hline
\textbf{Metric} & \textbf{Description} & \textbf{Formula} & \textbf{Theory} & \textbf{Cognitive Relevance} \\ \hline
CIB & Balances user interaction with automation. & \( \frac{\text{Number of Front-end Components}}{\text{Number of Back-end Interactions}} \) & Lumberjack model. Ensures neither under-engagement nor overwhelm. & Maintains engagement without overwhelming users. See Section \ref{sec:cib}. \\ \hline
\end{tabular}
\caption{Component Interaction Balance (CIB)}
\label{tab:CIB_tab}
\end{table}
The Component Interaction Balance (CIB) metric is designed to assess the ratio between front-end components (interfaces with which operators interact) and back-end interactions (automated system processes). This is in line with Brunswik's Lens Model and Gary Klein's Recognition-Primed Decision (RPD) model in Naturalistic Decision Making, where distal variables (environmental cues) should be accurately translated into proximal cues that inform judgment \cite{brunswik1955representative, Klein1993ARD}. In this context, front-end components should accurately convey relevant information to users to facilitate decision-making. The RPD model predicts that experts leverage patterns in data to make quick, effective decisions, so by evaluating the balance between the complexity of the front-end interface and the underlying back-end interactions, CIB aims to provide a measure of efficiency in decision-making \cite{Klein1993ARD}.

\begin{table}[H]
\tiny
\centering
\begin{tabular}{|>{\raggedright\arraybackslash}m{0.5cm}|>{\raggedright\arraybackslash}m{2cm}|>{\centering\arraybackslash}m{3cm}|>{\raggedright\arraybackslash}m{3cm}|>{\raggedright\arraybackslash}m{3cm}|}
\hline
\textbf{Metric} & \textbf{Description} & \textbf{Formula} & \textbf{Theory} & \textbf{Cognitive Relevance} \\ \hline
OP & Identifies excessive back-end demands. & \( \text{Max}(0, \text{Back-end} - \text{Front-end}) \) & Systemic automation failures. & Reduces user confusion and cognitive overload. See Section \ref{sec:op}. \\ \hline
\end{tabular}
\caption{Overload Penalty (OP)}
\label{tab:OP_tab}
\end{table}
The Overload Penalty (OP) metric aims at measuring the excess of back-end interactions compared to front-end components, identifying potential cognitive overload. This metric is grounded in Cognitive Load Theory, which posits that excessive cognitive load hinders learning and decision-making \cite{Sweller1988CognitiveLD}. The principle also follows from Wickens' Multiple Resource Theory (MRT), which states that humans have limited cognitive resources for processing information \cite{Wickens2002MultipleRA}. When back-end interactions outnumber the front-end components, it evidences a system complexity that can surpass the user's cognitive capacity. The OP metric seeks to evaluate and enable the minimisation of this discrepancy, promoting a design that aligns with human cognitive limitations.

\begin{table}[H]
\tiny
\centering
\begin{tabular}{|>{\raggedright\arraybackslash}m{0.5cm}|>{\raggedright\arraybackslash}m{2cm}|>{\centering\arraybackslash}m{3cm}|>{\raggedright\arraybackslash}m{3cm}|>{\raggedright\arraybackslash}m{3cm}|}
\hline
\textbf{Metric} & \textbf{Description} & \textbf{Formula} & \textbf{Theory} & \textbf{Cognitive Relevance} \\ \hline
ASE & Aligns system design with cognitive limits. & \( \frac{\text{Effective Components}}{\text{Total Components}} \) & Cognitive load theory and human working memory constraints. & Optimizes user interface for better cognitive processing. See Section \ref{sec:ase}. \\ \hline
\end{tabular}
\caption{Attention Span Efficiency (ASE)}
\label{tab:ASE_tab}
\end{table}
The Attention Span Efficiency (ASE) metric is designed to evaluate how well front-end components are chunked, inspired by Miller's Magical Number Seven theory and Cowan's chunking research. Miller famously suggested that humans can process up to seven plus or minus two chunks of information at a time, an hypothesis which was refined by Cowan by limiting effective working memory to four chunks, discrete pieces of information grouped together as a single unit based on familiarity or patterns \cite{miller1956magical, cowan2001magical}. ASE aims to quantify whether front-end components are grouped within this optimal range. ASE also aligns with the theory of policy capturing, where relevant cues should be presented in a format that users can interpret effectively \cite{Slovic1969AnalyzingTE}.

\begin{table}[H]
\tiny
\centering
\begin{tabular}{|>{\raggedright\arraybackslash}m{0.5cm}|>{\raggedright\arraybackslash}m{2cm}|>{\centering\arraybackslash}m{3cm}|>{\raggedright\arraybackslash}m{3cm}|>{\raggedright\arraybackslash}m{3cm}|}
\hline
\textbf{Metric} & \textbf{Description} & \textbf{Formula} & \textbf{Theory} & \textbf{Cognitive Relevance} \\ \hline
CSI & Measures cognitive demand by assessing task time and error rates. & \( \alpha \times \frac{\text{Task Time}}{\text{Base Time}} + \beta \times \text{Error Rate} \) & Theories of cognitive load and error management. & Identifies and reduces error rates and inefficiencies. See Section \ref{sec:csi}. \\ \hline
\end{tabular}
\caption{Cognitive Strain Index (CSI)}
\label{tab:CSI_tab}
\end{table}
The Cognitive Strain Index (CSI) attempts to measure task completion time and error rate as an evaluation of cognitive load. This metric is in line with Sweller's Cognitive Load Theory, where there is a need to manage intrinsic and extraneous cognitive load, where intrinsic cognitive load refers to the inherent complexity of the material engaging the operator and depends on prior knowledge, while extraneous cognitive load is the additional mental burden imposed by poor instructional design \cite{Sweller1988CognitiveLD}. Furthermore, our CSI design follows the principles of Signal Detection Theory, where error rates and decision-making accuracy are considered valuable indicators of cognitive strain \cite{Green1966SignalDT}.

\begin{table}[H]
\tiny
\centering
\begin{tabular}{|>{\raggedright\arraybackslash}m{1cm}|>{\raggedright\arraybackslash}m{2cm}|>{\centering\arraybackslash}m{4cm}|>{\raggedright\arraybackslash}m{2cm}|>{\raggedright\arraybackslash}m{2.5cm}|}
\hline
\textbf{Metric} & \textbf{Description} & \textbf{Formula} & \textbf{Theory} & \textbf{Cognitive Relevance} \\ \hline
Wasted Attention Resource (WAR) & Measures underutilization of cognitive capacity if the number of front-end components is fewer than five. & \( \max(0, 5 - \text{Number of Front-end Components}) \) & Cognitive load theory; ensures no cognitive resources are wasted. & Optimizes cognitive resource usage by maintaining minimum engagement levels. See Section \ref{sec:war}. \\ \hline
Noise Index (NI) & Quantifies excess cognitive load when front-end components exceed human processing capability. & \( \max(0, \text{Number of Front-end Components} - 9) \) & Cognitive load and information processing theories. & Prevents information overload by aligning with cognitive capabilities. See Section \ref{sec:ni}. \\ \hline
\end{tabular}
\caption{Wasted Attention Resource (WAR) and Noise Index (NI)}
\label{tab:NI_tab}
\end{table}
The Wasted Attention Resource (WAR) metric targets the underutilization of cognitive resources if fewer than five front-end components are present, following Miller's and Cowan's theories of chunking \cite{miller1956magical, cowan2001magical}. By defining a minimum engagement level, WAR encourages system designs that utilize available cognitive resources optimally, preventing attention underuse and enhancing situational awareness \cite{Endsley1995TowardAT}. This is supported further by the Naturalistic Decision Making (NDM) framework, where situational awareness is considered crucial for recognizing patterns and making decisions \cite{chapter4}. In contrast to WAR, the Noise Index (NI) metric quantifies excess cognitive load when front-end components exceed the average processing capability. It is derived from Cognitive Load Theory, which underscores the detrimental effects of information overload on performance \cite{Sweller1988CognitiveLD}. Moreover, Kahneman and Tversky's research on heuristics and biases indicates that humans rely on mental shortcuts when overwhelmed, leading to errors \cite{kahneman1982judgment}. The NI also follows from the Cognitive Continuum Theory, which suggests a spectrum between intuitive and analytical thinking, in that by reducing noise through thresholding chunking at 9 items, the NI metric aims for a balance between these cognitive modes so as to promote optimal decision-making \cite{Hammond1999JudgmentsUS}.

\begin{table}[H]
\tiny
\centering
\begin{tabular}{|>{\raggedright\arraybackslash}m{1cm}|>{\raggedright\arraybackslash}m{2cm}|>{\centering\arraybackslash}m{3cm}|>{\raggedright\arraybackslash}m{2cm}|>{\raggedright\arraybackslash}m{3cm}|}
\hline
\textbf{Metric} & \textbf{Description} & \textbf{Formula} & \textbf{Theory} & \textbf{Cognitive Relevance} \\ \hline
Interaction Redundancy (IR) & Evaluates unnecessary duplication in back-end processes. & \( \text{Total Back-end Interactions} - \text{Unique Back-end Interactions} \) & System efficiency and error reduction theories. & Reduces system complexity and potential user errors by minimizing redundant processes. See Section \ref{sec:ir}. \\ \hline
\end{tabular}
\caption{Interaction Redundancy (IR)}
\label{tab:IR_tab}
\end{table}
The Interaction Redundancy (IR) metric is designed to enable identification of unnecessary duplication in back-end processes, drawing on principles from Cognitive Load Theory and System Efficiency Theory, namely the minimisation of intrinsic and extraneous cognitive loads \cite{Sweller1988CognitiveLD, Green1966SignalDT}. By reducing redundant interactions, IR minimizes system complexity and potential user errors.
The IR metric aligns with the principles of Coherence Theory, as outlined by Hogarth and Einhorn, which emphasizes the importance of consistent and rational information processing \cite{Hogarth1992OrderEI}. By minimizing unnecessary duplication in back-end processes, the IR metric also ensures that users receive coherent and logical information, thereby reducing cognitive dissonance and enhancing their ability to make rational decisions \cite{chapter4}. Lastly, IR follows from Parasuraman's model of automation, where each level of automation is seen as contributing to overall system performance \cite{effects_imperfect}. Here, the aim is to do so without overwhelming users with redundant or irrelevant information. This alignment should help maintain optimal workload balance across automation levels, reducing user errors and enhancing the overall efficiency of human-automation interaction.

\begin{table}[H]
\tiny
\centering
\begin{tabular}{|>{\raggedright\arraybackslash}m{1cm}|>{\raggedright\arraybackslash}m{2cm}|>{\centering\arraybackslash}m{3cm}|>{\raggedright\arraybackslash}m{2cm}|>{\raggedright\arraybackslash}m{3cm}|}
\hline
\textbf{Metric} & \textbf{Description} & \textbf{Formula} & \textbf{Theory} & \textbf{Cognitive Relevance} \\ \hline
Operational Latency (OL) & Measures the delay between user action and system response. & \( t_{\text{response}} - t_{\text{action}} \) & System response time theory. & Ensures system responsiveness is aligned with user expectations and operational demands. See Section \ref{sec:ol}. \\ \hline
Critical Risk Index (CRI) & Assesses risk due to overlooked system interactions. & \( 2^{\max(0, \text{N Overlooked Interactions} - 9)} \) & Risk management and error prediction in critical systems. & Identifies high-risk scenarios potentially leading to catastrophic outcomes. See Section \ref{sec:cri}. \\ \hline
\end{tabular}
\caption{Operational Latency (OL) and Critical Risk Index (CRI)}
\label{tab:CRI_tab}
\end{table}
The Operational Latency (OL) metric assesses the delay between user action and system response, derived from principles in System Response Time Theory. Nielsen argues that a response time below one second maintains user engagement, while delays beyond ten seconds disrupt the cognitive flow \cite{Nielsen1993UsabilityE}. The metric also relates to Cognitive Continuum Theory, where intuitive decision-making relies on immediate feedback. By minimizing operational latency, systems can better support users in making intuitive decisions without disrupting their cognitive flow \cite{Hammond1999JudgmentsUS}.
The Critical Risk Index (CRI) quantifies the risk due to overlooked system interactions, drawing on Risk Management Theory and Signal Detection Theory, where in high-stakes environments, missing critical system interactions can lead to catastrophic outcomes \cite{Aven2016RiskAA, Green1966SignalDT}. According to the Coherence Model, overlooked signals often result from inconsistencies in information processing, therefore CRI is designed to identify scenarios where cognitive overload or system design flaws increase the likelihood of high-risk errors \cite{Hogarth1992OrderEI}.

\section{Discussion}
We think that the proposed benchmark framework and associated metrics encapsulated in $\Delta$-EVAL represent an initial yet significant step toward standardizing the evaluation of HAI systems in a holistic manner. One of the central challenges in this field is balancing the complexity of automated processes with the cognitive limitations of human operators - a challenge that has been well documented in prior research \cite{lee2013oxford, kahneman1982judgment, wickens2004introduction}. In $\Delta$-EVAL, metrics such as the Overload Penalty (OP) for identifying excessive back-end interactions and the Noise Index (NI) for detecting information overload at the front-end, are simple initial operationalizations aiming in the direction of providing quantitative insights into this balance.

Regardless - indeed, perhaps because of the simple nature of the current work - several limitations must be acknowledged. First, many of the metrics in $\Delta$-EVAL are based on simplified interpretations of complex cognitive phenomena. For example, using fixed chunking thresholds (e.g., fewer than five or more than nine front-end components) may not capture inter-individual variability or context-specific demands, as suggested by chunking theories \cite{cowan2001magical}. Second, while the metrics are theoretically motivated, they have yet to be validated through experimental studies \cite{effects_imperfect, karanikas}. Future work should focus on refining these measures based on empirical data and adapting them to a broad range of real-world tasks \cite{cnossen2004adaptive}.

Future work might also extend the current metrics by integrating machine learning techniques into $\Delta$-EVAL. By, for example, collecting time series data and user interaction logs, predictive models could be developed to forecast cognitive strain or attention lapses in real time \cite{kasneci2023chatgpt}. Such adaptive systems would allow for dynamic adjustments, thereby optimizing metrics like the Attention Span Efficiency (ASE) or Cognitive Strain Index (CSI) on-the-fly as well as drawing attention to potential systemic bottlenecks otherwise hidden from view which can have been perpetuated through, for example, UI design dogma. This synergy between classical evaluation metrics and data-driven models underscores the potential of $\Delta$-EVAL as a framework that can evolve alongside advancements in both AI benchmarking \cite{lm_evaluation_harness, srivastava2023imitation} and human factors research, as well as people themselves, as we increasingly interact with more diverse technological interfaces and generate data that can be fed back in the development loop.

\section{Conclusion}
In this work, we introduced $\Delta$-EVAL: a hypothetical benchmark framework for evaluating HAI systems, focusing on both front-end user interfaces and back-end automated processes. Drawing on established theories in Cognitive Engineering \cite{chapter4}, Human Factors \cite{wickens2004introduction} and broader Cognitive Science \cite{brunswik1955representative, kahneman1982judgment}, we proposed a suite of metrics ranging from Component Interaction Balance (CIB) and Attention Span Efficiency (ASE) to Operational Latency (OL) and Critical Risk Index (CRI) which represent small steps towards the quantitative evaluation of key aspects of the operator-system interaction.

Our approach seeks to unify diverse evaluation criteria into a single, coherent methodology. In doing so, $\Delta$-EVAL bridges the gap between traditional AI benchmarking methods \cite{lm_evaluation_harness, rajpurkar2016squad, wang2018glue, srivastava2023imitation, lintang_sutawika_2023_10256836} and the more nuanced requirements of human-centric system evaluation, as evidenced by previous studies on automation failures and decision-making processes \cite{Automation_failure, effects_imperfect, karanikas}. Although the proposed metrics are grounded in theoretical models - such as Brunswik’s Lens Model \cite{brunswik1955representative}, Naturalistic Decision Making \cite{effects_imperfect, sheridan2005human}, and Cognitive Continuum Theory \cite{sheridan2005human, kasneci2023chatgpt} - here they have been left largely conceptual. As such, empirical validation, task-specific calibration, and iterative refinement are essential next steps toward ensuring the practical applicability of HAI theory to the development of real-world applications through benchmark development and testing.

Ultimately, $\Delta$-EVAL is designed to try and provide a standardized, reproducible approach to assessing HAI systems. We believe that by integrating both technical and cognitive measures, our proposed benchmark can drive the development of automation systems that are not only efficient and reliable but also closely aligned with human cognitive capabilities - thereby enhancing safety and overall system performance in high-stakes environments.

In conclusion, we tried to show the promise and the challenges inherent in developing a unified benchmark for HAI systems. $\Delta$-EVAL not only promises a foundation for assessing the balance between front-end and back-end components but also offers a roadmap for future enhancements through iterative experimentation. As automation becomes increasingly integral to safety-critical applications, continued refinement and validation of such benchmarks and the underlying software infrastructure that supports HAI systems will be essential to ensure that human operators remain effectively engaged and supported.
\newpage

\appendix
\section{Appendix}
\subsection{Causal Inference Using Do-Calculus for HAI Benchmarks}\label{app:causal}
In this section we describe how to use Judea Pearl’s do-calculus and causal networks to quantify the causal effect of design interventions on cognitive outcomes in Human-Automation Interaction (HAI) systems. We assume that a causal model (i.e. a Directed Acyclic Graph, DAG) correctly represents the assumptions underlying the relationships between the intervention variable, cognitive outcomes, and any confounders.

\subsubsection{Preliminaries and Notation}
Let 
\begin{itemize}
  \item $X$ denote the intervention (e.g. a front-end design parameter),
  \item $Y$ denote the cognitive outcome (e.g. cognitive load, decision accuracy),
  \item $Z$ denote a set of observed confounders (e.g. user expertise, baseline stress) that affect both $X$ and $Y$.
\end{itemize}

The causal effect of setting $X=x$ is given by the interventional distribution
\[
P(Y \mid do(X=x)).
\]
Under the assumptions encoded in the DAG, our goal is to identify $P(Y\mid do(X=x))$ from observational data.

\subsubsection{Identification via the Back-Door Criterion}
If there exists a set of variables $Z$ that blocks every back-door path from $X$ to $Y$, then by the back-door adjustment we have
\begin{equation}
P(Y \mid do(X=x)) = \sum_{z} P(Y \mid X=x, Z=z) P(Z=z).
\label{eq:backdoor}
\end{equation}
For example, if $Z$ captures all common causes of $X$ and $Y$, then the above formula identifies the causal effect of $X$ on $Y$.

\textbf{Do-Calculus Rules}\\
Pearl’s do-calculus provides three rules for transforming expressions with interventions. For instance, the first rule states that if $Y$ is independent of $Z$ given $X$ and $W$ in the mutilated graph (denoted $G_{\overline{X}}$), then
\begin{equation}
P(Y \mid do(X), Z, W) = P(Y \mid do(X), W).
\label{eq:rule1}
\end{equation}
Rules of this type allow one to remove or insert observations or actions provided the conditional independence relationships hold. These rules are essential when more complex causal structures (including mediators and colliders) are involved.

\subsubsection{Causal Mediation Analysis}
In many HAI systems an intervention $X$ may affect $Y$ both directly and indirectly through a mediator $M$ (for example, a change in interface design might alter user attention $M$, which in turn affects cognitive performance $Y$). The total effect (TE) can be decomposed into:
\begin{align}
\text{NDE} &= E[Y \mid do(X=1, M=M_{x=0})] - E[Y \mid do(X=0)], \\
\text{NIE} &= E[Y \mid do(X=1)] - E[Y \mid do(X=1, M=M_{x=0})],
\end{align}
where $M_{x=0}$ is the potential mediator value when $X=0$. The Average Treatment Effect (ATE) is then
\begin{equation}
ATE = E[Y \mid do(X=1)] - E[Y \mid do(X=0)].
\label{eq:ate}
\end{equation}

\subsubsection{A Simple Causal Diagram}
Figure~\ref{fig:causalDAG} shows a simple causal graph in which $Z$ confounds the relationship between $X$ and $Y$. Conditioning on $Z$ (i.e. using Eq.~\eqref{eq:backdoor}) removes the spurious association.
\begin{figure}[H]
\centering
\begin{tikzpicture}[->,>=stealth, node distance=2.5cm, semithick]
    \node (Z) [draw, circle] {$Z$};
    \node (X) [draw, circle, right of=Z, xshift=2cm] {$X$};
    \node (Y) [draw, circle, right of=X, xshift=2cm] {$Y$};
    \draw (Z) edge (X);
    \draw (Z) edge (Y);
    \draw (X) edge (Y);
\end{tikzpicture}
\caption{A DAG where $Z$ is a confounder for the effect of intervention $X$ on cognitive outcome $Y$.}
\label{fig:causalDAG}
\end{figure}

Figure~\ref{fig:mediatorDAG} illustrates a scenario with a mediator $M$, which allows for the decomposition of the total effect into direct and indirect effects.
\begin{figure}[H]
\centering
\begin{tikzpicture}[->,>=stealth, node distance=2.5cm, semithick]
    \node (X) [draw, circle] {$X$};
    \node (M) [draw, circle, right of=X, xshift=2cm] {$M$};
    \node (Y) [draw, circle, right of=M, xshift=2cm] {$Y$};
    \draw (X) edge (M);
    \draw (M) edge (Y);
    \draw (X) edge[bend left] (Y);
\end{tikzpicture}
\caption{A DAG with a mediator $M$, permitting causal mediation analysis of the effect of $X$ on $Y$.}
\label{fig:mediatorDAG}
\end{figure}





\newpage



  
\bibliographystyle{plain}
\bibliography{references}

\begin{thebibliography}{10}

\bibitem{arulkumaran2017deep}
Kai Arulkumaran, Marc~Peter Deisenroth, Miles Brundage, and Anil~Anthony Bharath.
\newblock Deep reinforcement learning: A brief survey.
\newblock {\em IEEE Signal Processing Magazine}, 34(6):26--38, 2017.

\bibitem{Aven2016RiskAA}
Terje Aven.
\newblock Risk assessment and risk management: Review of recent advances on their foundation.
\newblock {\em Eur. J. Oper. Res.}, 253:1--13, 2016.

\bibitem{brunswik1955representative}
Egon Brunswik.
\newblock Representative design and probabilistic theory in a functional psychology.
\newblock {\em Psychological review}, 62(3):193, 1955.

\bibitem{cnossen2004adaptive}
Fokie Cnossen, Theo Meijman, and Talib Rothengatter.
\newblock Adaptive strategy changes as a function of task demands: A study of car drivers.
\newblock {\em Ergonomics}, 47:218--36, 03 2004.

\bibitem{cowan2001magical}
Nelson Cowan.
\newblock The magical number 4 in short-term memory: A reconsideration of mental storage capacity.
\newblock {\em Behavioral and brain sciences}, 24(1):87--114, 2001.

\bibitem{crocoll1990status}
William~M Crocoll and Bruce~G Coury.
\newblock Status or recommendation: Selecting the type of information for decision aiding.
\newblock {\em Unknown}, 34:1524--1528, 1990.

\bibitem{deng2009imagenet}
Jia Deng, Wei Dong, Richard Socher, Li-Jia Li, Kai Li, and Li~Fei-Fei.
\newblock Imagenet: A large-scale hierarchical image database.
\newblock {\em Unknown}, pages 248--255, 2009.

\bibitem{lm_evaluation_harness}
EleutherAI.
\newblock Language model evaluation harness.
\newblock \url{https://github.com/EleutherAI/lm-evaluation-harness/tree/main/lm_eval/tasks}, 2023.

\bibitem{Endsley1995TowardAT}
Mica~R. Endsley.
\newblock Toward a theory of situation awareness in dynamic systems.
\newblock {\em Human Factors: The Journal of Human Factors and Ergonomics Society}, 37:32 -- 64, 1995.

\bibitem{endsley}
Mica~R. Endsley.
\newblock The divergence of objective and subjective situation awareness: A meta-analysis.
\newblock {\em Journal of Cognitive Engineering and Decision Making}, 14(1):34--53, 2020.

\bibitem{Erdil2023ExplosiveGF}
Ege Erdil and Tamay Besiroglu.
\newblock Explosive growth from ai automation: A review of the arguments.
\newblock {\em Unknown}, 2023.

\bibitem{Green1966SignalDT}
David~M. Green and John~A. Swets.
\newblock Signal detection theory and psychophysics.
\newblock {\em Unknown}, 1966.

\bibitem{Automation_failure}
Jr. Gyrd~Skraaning and Greg~A. Jamieson.
\newblock The failure to grasp automation failure.
\newblock {\em Journal of Cognitive Engineering and Decision Making}, 0(0):15553434231189375, 0.

\bibitem{Hammond1999JudgmentsUS}
Kenneth~R. Hammond.
\newblock Judgments under stress.
\newblock {\em Unknown}, 1999.

\bibitem{Hogarth1992OrderEI}
Robin~M. Hogarth and Hillel~J. Einhorn.
\newblock Order effects in belief updating: The belief-adjustment model.
\newblock {\em Cognitive Psychology}, 24:1--55, 1992.

\bibitem{kahneman1982judgment}
Daniel Kahneman, Paul Slovic, and Amos Tversky.
\newblock {\em Judgment under uncertainty: Heuristics and biases}.
\newblock Cambridge university press, 1982.

\bibitem{karanikas}
Nektarios Karanikas.
\newblock Human error views: A framework for benchmarking organizations and measuring the distance between academia and industry.
\newblock {\em Unknown}, 10 2015.

\bibitem{kasneci2023chatgpt}
Enkelejda Kasneci, Kathrin Se{\ss}ler, Stefan K{\"u}chemann, Maria Bannert, Daryna Dementieva, Frank Fischer, Urs Gasser, Georg Groh, Stephan G{\"u}nnemann, Eyke H{\"u}llermeier, et~al.
\newblock Chatgpt for good? on opportunities and challenges of large language models for education.
\newblock {\em Learning and individual differences}, 103:102274, 2023.

\bibitem{Klein1993ARD}
Gary Klein.
\newblock A recognition-primed decision (rpd) model of rapid decision making.
\newblock {\em Unknown}, 1993.

\bibitem{lee2013oxford}
John~D Lee and Alex Kirlik.
\newblock {\em The Oxford handbook of cognitive engineering}.
\newblock Oxford University Press, 2013.

\bibitem{lin2014microsoft}
Tsung-Yi Lin, Michael Maire, Serge Belongie, James Hays, Pietro Perona, Deva Ramanan, Piotr Doll{\'a}r, and C~Lawrence Zitnick.
\newblock Microsoft coco: Common objects in context.
\newblock {\em Unknown}, pages 740--755, 2014.

\bibitem{miller1956magical}
George~A Miller.
\newblock The magical number seven, plus or minus two: Some limits on our capacity for processing information.
\newblock {\em Psychological review}, 63(2):81, 1956.

\bibitem{mnih2013playing}
Volodymyr Mnih, Koray Kavukcuoglu, David Silver, Alex Graves, Ioannis Antonoglou, Daan Wierstra, and Martin Riedmiller.
\newblock Playing atari with deep reinforcement learning.
\newblock {\em arXiv preprint arXiv:1312.5602}, 2013.

\bibitem{chapter4}
Kathleen Mosier.
\newblock {68 Judgment and Prediction}.
\newblock In {\em {The Oxford Handbook of Cognitive Engineering}}. Oxford University Press, 02 2013.

\bibitem{Nielsen1993UsabilityE}
Jakob Nielsen.
\newblock Usability engineering.
\newblock {\em Unknown}, page Unknwon, 1993.

\bibitem{onnasch2014human}
Linda Onnasch, Christopher~D Wickens, Huiyang Li, and Dietrich Manzey.
\newblock Human performance consequences of stages and levels of automation: An integrated meta-analysis.
\newblock {\em Human factors}, 56(3):476--488, 2014.

\bibitem{rajpurkar2016squad}
Pranav Rajpurkar, Jian Zhang, Konstantin Lopyrev, and Percy Liang.
\newblock Squad: 100,000+ questions for machine comprehension of text.
\newblock {\em arXiv preprint arXiv:1606.05250}, 2016.

\bibitem{effects_imperfect}
Ericka Rovira, Kathleen Mcgarry, and Raja Parasuraman.
\newblock Effects of imperfect automation on decision making in a simulated command and control task.
\newblock {\em Human factors}, 49:76--87, 03 2007.

\bibitem{sarter2001supporting}
Nadine~B Sarter and Beth Schroeder.
\newblock Supporting decision making and action selection under time pressure and uncertainty: The case of in-flight icing.
\newblock {\em Human factors}, 43(4):573--583, 2001.

\bibitem{sattler2019understanding}
Torsten Sattler, Qunjie Zhou, Marc Pollefeys, and Laura Leal-Taixe.
\newblock Understanding the limitations of cnn-based absolute camera pose regression.
\newblock {\em Unknown}, pages 3302--3312, 2019.

\bibitem{sebok2017implementing}
Angelia Sebok and Christopher~D Wickens.
\newblock Implementing lumberjacks and black swans into model-based tools to support human--automation interaction.
\newblock {\em Human factors}, 59(2):189--203, 2017.

\bibitem{sheridan2005human}
Thomas~B Sheridan and Raja Parasuraman.
\newblock Human-automation interaction.
\newblock {\em Reviews of human factors and ergonomics}, 1(1):89--129, 2005.

\bibitem{silver2016mastering}
David Silver, Aja Huang, Chris~J Maddison, Arthur Guez, Laurent Sifre, George Van Den~Driessche, Julian Schrittwieser, Ioannis Antonoglou, Veda Panneershelvam, Marc Lanctot, et~al.
\newblock Mastering the game of go with deep neural networks and tree search.
\newblock {\em nature}, 529(7587):484--489, 2016.

\bibitem{Slovic1969AnalyzingTE}
Paul Slovic.
\newblock Analyzing the expert judge: A descriptive study of a stockbroker's decision process.
\newblock {\em Journal of Applied Psychology}, 53:255--263, 1969.

\bibitem{sokolova2009systematic}
Marina Sokolova and Guy Lapalme.
\newblock A systematic analysis of performance measures for classification tasks.
\newblock {\em Information processing \& management}, 45(4):427--437, 2009.

\bibitem{srivastava2023imitation}
Aarohi Srivastava and (others).
\newblock Beyond the imitation game: Quantifying and extrapolating the capabilities of language models.
\newblock {\em Unknown}, 2023.

\bibitem{stowers}
Kimberly Stowers, James Oglesby, Shirley Sonesh, Kevin Leyva, Chelsea Iwig, and Eduardo Salas.
\newblock A framework to guide the assessment of human–machine systems.
\newblock {\em Human Factors}, 59:172--188, 03 2017.

\bibitem{lintang_sutawika_2023_10256836}
Lintang Sutawika et~al.
\newblock Eleutherai/lm-evaluation-harness: Major refactor.
\newblock {\em Unknown}, dec 2023.

\bibitem{Sweller1988CognitiveLD}
John Sweller.
\newblock Cognitive load during problem solving: Effects on learning.
\newblock {\em Cogn. Sci.}, 12:257--285, 1988.

\bibitem{wang2018glue}
Alex Wang, Amanpreet Singh, Julian Michael, Felix Hill, Omer Levy, and Samuel~R Bowman.
\newblock Glue: A multi-task benchmark and analysis platform for natural language understanding.
\newblock {\em arXiv preprint arXiv:1804.07461}, 2018.

\bibitem{Wickens2002MultipleRA}
Christopher~D. Wickens.
\newblock Multiple resources and performance prediction.
\newblock {\em Theoretical Issues in Ergonomics Science}, 3:159 -- 177, 2002.

\bibitem{wickens2004introduction}
Christopher~D Wickens, Sallie~E Gordon, Yili Liu, and J~Lee.
\newblock {\em An introduction to human factors engineering}, volume~2.
\newblock Pearson Prentice Hall Upper Saddle River, NJ, 2004.

\bibitem{yeh2003head}
Michelle Yeh, James~L Merlo, Christopher~D Wickens, and David~L Brandenburg.
\newblock Head up versus head down: The costs of imprecision, unreliability, and visual clutter on cue effectiveness for display signaling.
\newblock {\em Human factors}, 45(3):390--407, 2003.

\end{thebibliography}

\end{document}